\documentclass[reqno,a4paper,12pt]{amsart}
\usepackage{subcaption}
\usepackage{amsmath}
\usepackage{amsfonts}
\usepackage{amssymb}
\usepackage{bm}
\usepackage{color}
\usepackage{lipsum}
\usepackage{graphicx}
\usepackage{epstopdf}

\usepackage[colorlinks=true,breaklinks=true,linkcolor=lightgreen,citecolor=lightblue]{hyperref}

\definecolor{lightblue}{rgb}{0.22,0.45,0.70}
\definecolor{cgray}{rgb}{0.7,0.7,0.7}

\definecolor{lightgreen}{rgb}{0.22,0.55,0.20}

\numberwithin{figure}{section}
\numberwithin{table}{section}
\numberwithin{equation}{section}

\DeclareMathAlphabet\mathbit
    \encodingdefault\rmdefault\bfdefault\itdefault
\DeclareOldFontCommand{\bi}{\normalfont\bfseries\itshape}{\mathbit}

\renewcommand{\geq}{\geqslant}
\renewcommand{\leq}{\leqslant}

\newcommand{\upd}{\mathrm{d}}
\newcommand{\rp}{\textcolor{black}}

\newcommand{\Lop}{{\mathcal N}}

\newcommand{\cH}{{\mathcal H}}
\newcommand{\cL}{{\mathcal L}}

\newcommand \beq{\begin{equation}}
\newcommand \eeq{\end{equation}}

\allowdisplaybreaks 

\begin{document}

\title[{Buckling Circumferential Ridge}] {Axisymmetric Ridges and Circumferential Buckling of Indented Shells of Revolution}

\author[M. Taffetani]{Matteo Taffetani}
\address{Department of Engineering Mathematics, University of Bristol, Ada Lovelace Building, University Walk, Bristol, BS8 1TW}
\author[M. G. Walker]{Martin G. Walker}
\address{Department of Civil and Environmental Engineering, University of Surrey, Guildford, Surrey, GU2 7XH}

\date{\today}

\maketitle


\begin{abstract} 
		When poking a thin shell-like structure, like a plastic water bottle, experience shows that an initial axisymmetric dimple forms around the indentation point. The ridge of this dimple, with increasing indentation, eventually buckles into a polygonal shape. The polygon order generally continues to increase with further indentation. In the case of spherical shells, both the underlying axisymmetric deformation and the buckling evolution have been studied in detail. However, little is known about the behaviour of general geometries. 
		
		In this work we describe the geometrical and mechanical features of the axisymmetric ridge that forms in indented general shells of revolution with non-negative Gaussian curvature and the conditions for circumferential buckling of this ridge. We show that, under the assumption of `mirror buckling' a single unified description of this ridge can be written if the problem is non-dimensionalised using the local slope of the undeformed shell mid-profile at the ridge radial location. However, in dimensional form the ridge properties evolve in quite different ways for different mid-profiles. Focusing on the indentation of shallow shells of revolution with constant Gaussian curvature, we use our theoretical framework to study the properties of the ridge at the circumferential buckling threshold and evaluate the validity of the mirror buckling assumption against a linear stability analysis on the shallow shell equations, showing very good agreement. Our results highlight that circumferential buckling in indented thin shells is controlled by a complex interplay between the geometry and the stress state in the ridge. 
		
		The results of our study will provide greater insight into the mechanics of thin shells. This could enable indentation to be used as a means to measure the mechanical properties of a wide range of shell geometries or used to design shells with specific mechanical behaviours.
	\end{abstract}

	\section{Introduction}
	
	The simple act of indenting a thin shell exposes a rich phenomenology that reveals the  principles underpinning the mechanics of these structures. Beyond the everyday examples of poking ping-pong ball or a plastic water bottle, the mechanics governing the indentation of thin shells applies across all length scales ranging from cell walls to aircraft fuselages. Indentation can also be used to non-destructively measure the mechanical properties of an object, such as biological cells \cite{Vella2012}. 
	
	In the analysis of indented thin shells, a great deal of study has been devoted to investigating the indentation of pressurised \cite{Marthelot2017a, Taffetani2017, Vella2012, Vella2011} and unpressurised  \cite{Knoche2014, Nasto2013, Nasto2014, Pauchard1998, Gupta2008, Taffetani2018} spherical shells. In both cases, the initial deformed shape is a localised dimple around the indentation point. Under increasing indentation, the dimple eventually loses axisymmetry. Pressurised shells buckle into a wrinkled shape, while the dimple ridge of unpressurised shells buckles into a polygonal shape.
	
	Analysis of the dimple ridge requires a description of the initial shape of the dimpled shell. Pogorelov \cite{Pogorelov1988} proposed an energetically \emph{cheap} axisymmetric deformation called `mirror buckling' where the deformed (dimpled) region is a mirror image of the corresponding region of the undeformed shell reflected about the plane of the dimple ridge. Under this assumption, the dimple is unstretched and the elastic strain energy is concentrated in the ridge region joining the deformed and undeformed regions of the shell. Despite the prominence of this assumption, mirror buckling has only been demonstrated for spherical shells undergoing de-pressurisation \cite{Knoche2014}. For indented unpressurised spherical shells the behaviour is more complex due to the stress singularity in the region of the indenter \cite{Gomez2016}. Seffen \cite{Seffen2016a} derived expressions for the axisymmetric deformed shape and mechanics of a conical shell undergoing inversion by employing the `mirror buckling' assumption which showed good agreement with finite element simulations.
	
	The geometry of a shell clearly influences its deformation response but, beyond the extensive research on spherical shells, the relationship between indentation and geometry has received little attention. Vaziri and Mahadevan \cite{Vaziri2008} discussed the influence of Gaussian curvature and show that, in shells with positive Gaussian curvature, the deformations are localised in the neighborhood of the indentation, while shells with negative Gaussian curvature exhibit non-local deformations which extend to the shell boundary. For shells with zero Gaussian curvature, the deformation is strongly anisotropic and influenced by the indentation location and boundary conditions \cite{Vella2012, Vaziri2008, Boudaoud2000}. The response to indentation of ellipsoidal and cylindrical shells (the latter obtained as the limit of ellipsoidal shells with one radius of curvature going to infinity) reveals the role of mean curvature in controlling the indentation stiffness, and Gaussian curvature in defining the influence of the boundary conditions on its response \cite{Lazarus2012,Vella2012,Sun2021}. 
	
	For cases of unpressurised shells with non-negative Gaussian curvature, the dimple ridge around the indentation point eventually undergoes circumferential buckling (sometimes called `secondary buckling') into a polygonal shape. Evaluation of the circumferential buckling threshold of the dimple ridge in unpressurised spherical shells has been investigated considering the effect of the boundary conditions, shallowness of the shell, and how these affect the order of the polygonal post-buckled ridge shape of the ridge \cite{Fitch1968,Taffetani2018}. However, the role of the shell shape on this buckling phenomenon is not clear. In the post-buckling regime it is well known that increasing indentation causes the ridge to progress through sequentially higher polygon orders \cite{Pauchard1998,Vaziri2008,Nasto2013}, although a satisfactory analytical model of this phenomenon remains elusive.
	
	\begin{figure*}
		\centering
		\includegraphics[width=0.8\textwidth]{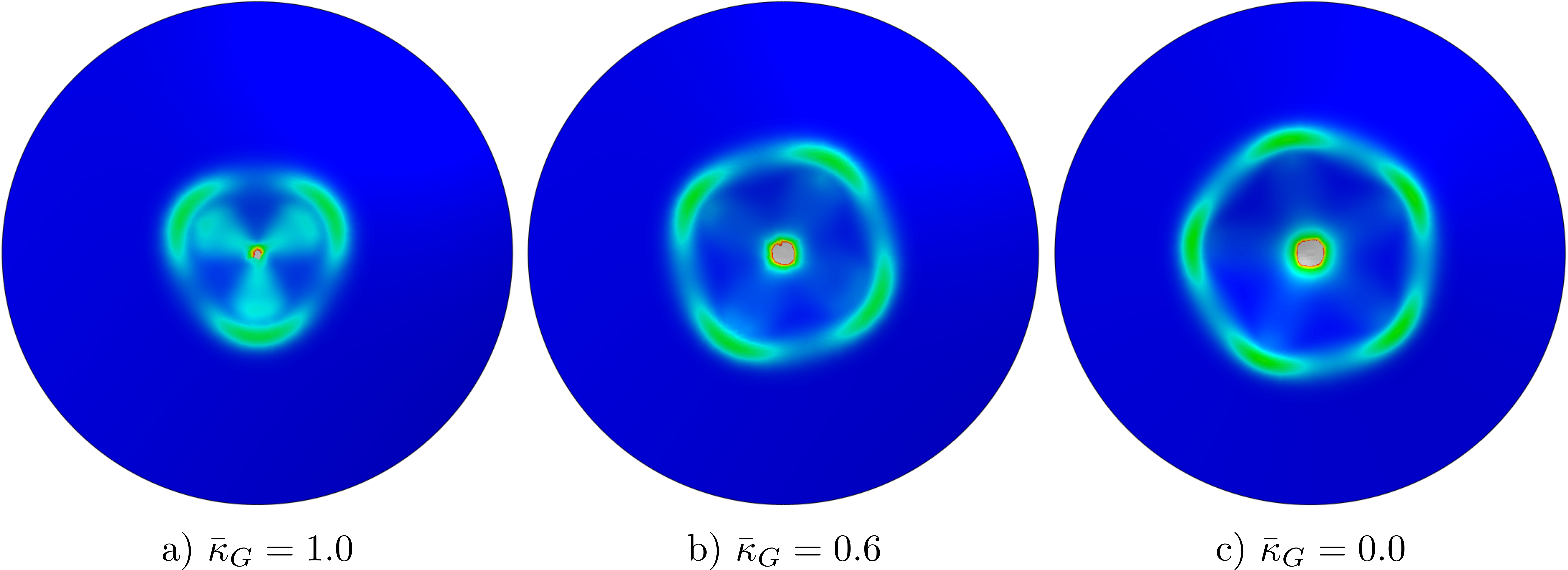}
		\caption{Deformed shapes (showing strain energy density contours) for the initial circumferential buckling of shells of revolution undergoing apex indentation with $\lambda=25$ and constant dimensionless Gaussian curvatures a) $\bar{\kappa}_G = 1.0$ (sphere), b) $\bar{\kappa}_G = 0.6$, and c) $\bar{\kappa}_G = 0.0$ (cone) obtained from finite element analysis. Definitions of $\lambda$ and $\bar{\kappa}_G$ are provided in Sections~\ref{sec:shallowShell} and \ref{SEC:circumBuckling} respectively. The circumferential buckling mode is clearly influenced by the shape of the indented shell.}
		\label{fig:KGRenderings}
	\end{figure*}
	
	In this paper we propose a unified description for the formation, and circumferential buckling, of the axisymmetric ridge of shallow shells of revolution with non-negative Gaussian curvature under point indentation at the apex. \rp{We choose shells with non-negative Gaussian curvatures since indention of shells with negative Gaussian curvature does not form the localised features which are the subject of this study \cite{Vaziri2008}.}  Employing the approach proposed by Pogorelov \cite{Pogorelov1988} for the description of the circular ridge in spherical shells undergoing mirror buckling, and adapting the formalism presented by Kierfield and Knoche \cite{Knoche2014}, we extend this description to a generic shallow shell of revolution. We show that a general description can be written if the problem is rescaled using the local slope of the undeformed mid-profile at the radial position of ridge. 
	
	By performing a near threshold analysis on the shallow shell equations, we obtain the expected circumferential buckling mode and critical indentation when the ridge buckles. With this description, we can predict the geometrical and mechanical features at the circumferential buckling threshold of the circular ridge in a general shell of revolution. Figure~\ref{fig:KGRenderings} shows the strain energy density contours of the deformed shape after circumferential buckling of shells with constant Gaussian curvature ranging between a cone and a sphere.  The shape of the shell clearly influences the circumferential buckling phenomenon with the mode number ranging between 3 (sphere) and 5 (cone).
	
	To evaluate our description we perform a linear stability analysis on the shallow shell equations and compare the results to our theoretical description. The results are also validated against finite element analysis showing very good agreement. The results of this study provide insight into the role of the shell geometry on the properties of the axisymmetric ridge of indented shells of revolution and subsequent circumferential buckling.

\section{The Pogorelov Ridge in Mirror Buckled Shells of Revolution}\label{SEC:PogorelovCone}

	We begin our investigation by considering the ridge of the axisymmetric dimple formed when a shell with non-negative Gaussian curvature is indented. We establish a global cylindrical coordinate system with radial coordinate $r$, azimuthal coordinate, $\theta$, and axial coordinate $z$. The base of a shell of revolution lies in the $(r,\theta)$ plane with $0<r\leq L$, where $L$ is the outer radius of the projection of the shell onto the $(r,\theta)$ plane. A generic shell of revolution is defined via the undeformed mid-profile $z_0(r) \geq 0$, with $z_0(0)$ being the $z$ coordinate of the apex of the shell.  We restrict our analysis to shallow shells with non-negative Gaussian curvature, such that $\kappa_{r}\kappa_{\theta}\geq 0$ with  $\kappa_{r} =  z_0''$ and $\kappa_{\theta} = z_0'/r $, where $\kappa_r$ and $\kappa_\theta$ are the radial and azimuthal curvatures respectively and the prime indicates the derivative with respect to the argument of the function. We assign the shell a uniform thickness, $t$, and a linear elastic material with Poisson's ratio, $\nu$, and Young's modulus, $E$. 
	
\begin{figure}[!t]
		\centering
		\includegraphics[width=0.5\textwidth]{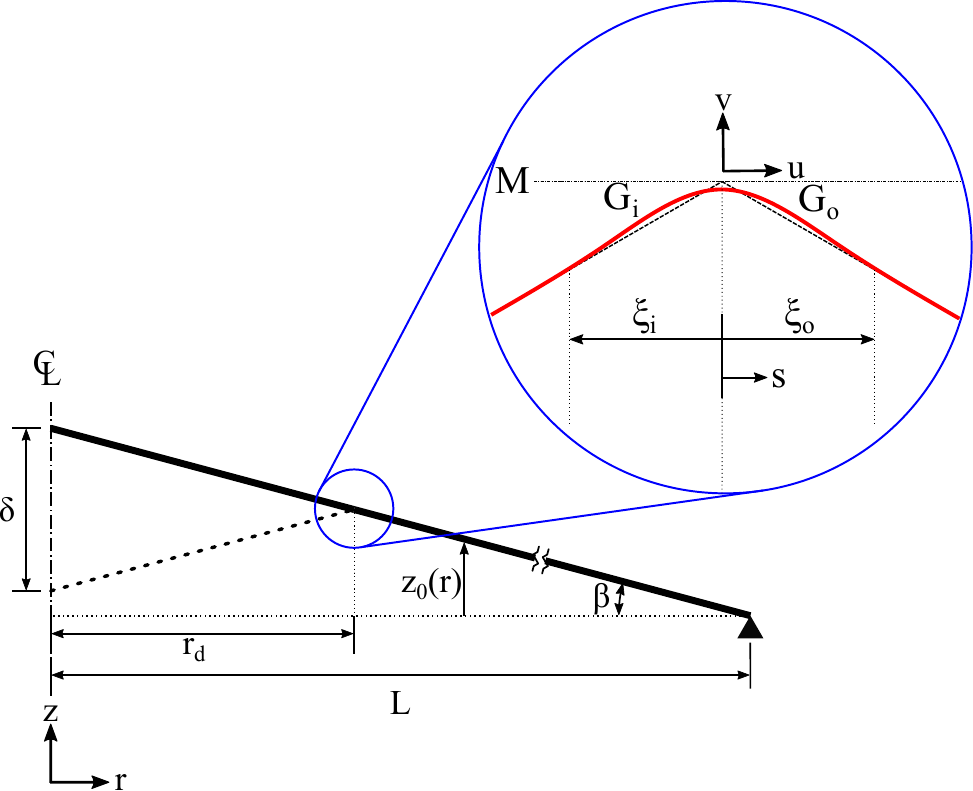}
		\caption{Schematic cross-section of a shallow shell in the process of inversion for the case of a conical shell of base angle $\beta$. The inset shows the ridge geometry, together with the local reference system used to describe the local deformation of the ridge. \rp{The line labeled M in the inset shows the mirror line for mirror buckling.}}\label{fig:schematic}
	\end{figure}

	We study the situation where the apex of the shell is displaced in the negative $z$ direction by a distance $\delta$ while the base of the shell remains on the $(r,\theta)$ plane. An axisymmetric dimple forms about the indentation point with a ridge located some projected distance, $r_d$, from the axis of revolution. Pogorelov \cite{Pogorelov1988} assumes that the deformed region of the shell, within the ridge, is the mirror image of the corresponding region of the undeformed shell reflected about the plane of the ridge, as shown in Fig.~\ref{fig:schematic} for a conical geometry.  In the limit of zero thickness, this `mirror buckled' region is an isometric deformation where the region of the shell within the ridge is turned `inside out' (thus changing the sign of the curvature while preserving lengths) and glued to the undeformed outer region. When the thickness is small but finite, the circular line that connects the two regions turns into a ridge whose geometry is set by the balance of the local stretching and bending energies \cite{Witten2007,Lobkovsky1996a}. Pogorelov assumes that the ridge can be described by a displacement field superimposed onto the `mirror buckled' base deformation. This has two major advantages: the energetic cost of this base deformation involves only the bending contribution, thus simplifying the expression of the strain energy; and it is possible to univocally compute the radial position of the ridge, $r_d$, once the displacement, $\delta$, of the apex is given and vice versa, although not always in closed form.
	
	To obtain insight into the role of the undeformed mid-profile shape on the axisymmetric ridge of an indented shell of revolution, we adapt the approach that Pogorelov \cite{Pogorelov1988} and  Knoche and Kierfeld \cite{Knoche2014} employed for the analysis of spherical shells. The reader is therefore directed to these references for full calculation details. Here we present aspects that are important for our shells of revolution. 
	
\subsection{Description of the ridge}\label{sec:ridgeStructure}
For now we assume that our shells of revolution with an undeformed mid-profile, $z_0(r)$, deform in a mirror buckled form, regardless of how this deformation is physically obtained. We examine this assumption in Section \ref{sec:results} for the case of indentation of the apex. The mirror buckled deformation of our shells is described by dividing them into three parts (neglecting the complexity discussed in \cite{Gomez2016}): the `mirror buckled' inner region, the undeformed outer region, and the ridge region. In the inner region, lengths are preserved and the strain energy has only the bending contribution due to the inversion of the shell curvature. The ridge region is centered at the coordinate $r=r_d$, which describes the position of the ridge apex in the limit of perfect mirror buckling and zero thickness. The ridge spans the radial region $\left[r_d-\xi_i,r_d+\xi_o\right]$ (see inset of Fig.~\ref{fig:schematic}).

Introducing a new coordinate $s=r-r_d$, the ridge can be further divided into two regions: an internal region $G_i = \{s:\left[-\xi_i, 0\right]\}$ and an external region $G_o = \{s:\left[0,\xi_o\right]\}$. Thus $s=0$ is the position of the ridge in this local reference system. Displacements with respect to the `mirror buckled' solution are denoted by $u(s)$ and $v(s)$, in the radial (since we focus on shallow geometries) and axial directions respectively. Under the assumption that the meridional strain vanishes, i.e. $\epsilon_r \approx 0$, the stretching energy depends only on the circumferential strain $\epsilon_{\theta} = u(s)/(s+r_d)$, where we can neglect the term $v \kappa_{r}$ because of the shallow assumption. The bending energy is computed from the change in curvature from the undeformed configuration. In the external region, $G_o$, the change in curvatures are due only to the local displacement field of the ridge and we can write $K_r (G_o)= v''(s)$ and $K_{\theta} (G_o) = v'(s)/(s+r_d)$. In the internal region, $G_i$, the effect of the local displacement field is superimposed onto the underlying global displacement so that, under the mirror buckling assumption, the change in meridional and circumferential curvatures read $K_r (G_i)= -2 \kappa_{r}  +v''(s) = - 2 z_0'' +v''(s)$ and $K_{\theta} (G_i)= -2 \kappa_{\theta} +v'(s)/(s+r_d)=  - 2 z_0'/(s+r_d) +v'(s)/(s+r_d)$ respectively. The strain energy due to stretching and bending, respectively, can then be written as
\begin{align}
	&U_s = \frac{1}{2} \frac{E h}{1-\nu^2}\int_G\left(\epsilon_r^2 + 2 \nu \epsilon_r \epsilon_\theta +\epsilon_\theta^2\right) \upd G, \\ &U_b = \frac{E h^3}{12\left(1-\nu^2\right)}\int_G\left(K_r^2+2\nu K_r K_\theta + K_\theta^2\right)\upd G 
\end{align}

To compute the unknown displacements, $u$ and $v$, one can use the equilibrium equations obtained by taking the first variation of the energy functional $U=U_s+U_b$. After inserting the expressions for the strains and changes in curvature, this functional can be further simplified assuming the conditions $v'(\pm \xi_o) = v'(\pm \xi_i)=0$ and $v'(0) = -z'_0(0)$, with the latter ensuring that the ridge has zero slope at its apex. Moreover, since the ridge is assumed to be shallow, $v'(r) \ll 1$. We additionally assume the ridge is narrow such that $\xi_i, \xi_o \ll 1$, and we can approximate $s+r_d\approx r_d$ and the differential area $\upd G \approx 2 \pi r_d \,\upd s$. 

Neglecting constant terms that do not enter in the variational problem over $u$ and $v$, the energy functional to minimize reads
\begin{equation}\label{EQ:EnergyKN}\begin{split}
& U = U_s(G_i) + U_b(G_i) +  U_s(G_o) + U_b(G_o) =\\
& \frac{E h}{(1-\nu^2)}\frac{\pi}{r_d}\int_{-\xi_i}^0 u^2 \;\upd s + \frac{E h^3}{12(1-\nu^2)} \pi r_d \int_{-\xi_i}^0 \left(v''\right)^2 \upd s +\\
& \frac{E h}{(1-\nu^2)}\frac{\pi}{r_d}\int_0^{\xi_o} u^2 \;\upd s + \frac{E h^3}{12(1-\nu^2)} \pi r_d \int_0^{\xi_o} \left(v''\right)^2 \upd s +\\
& \frac{E h^3}{12(1-\nu^2)} \pi r_d\int_{-\xi_i}^0 \left[-4 v''\left(\kappa_{r} + \nu \kappa_{\theta}\right) - 4 \frac{v'}{r_d}\left(\nu \kappa_{r} + \kappa_{\theta}\right)\right] \upd r
\end{split}\end{equation}
The assumption that $\epsilon_r = 0$ leads to the constraint equation \cite{Pogorelov1988}:
\begin{equation}
u'(s) + \frac{1}{2}\left(v'(s)\right)^2 + |z_0'(r_d)| v'(s) = 0, \label{EQ:epsilons0}
\end{equation}
where we can see the local slope of the undeformed shape at the ridge location, $z_0'(r_d)$, entering explicitly in the governing equations of the ridge.

The energy functional in Eqn~\eqref{EQ:EnergyKN} differs from what was derived in \cite{Knoche2014} by the term in the last row, where we recall that $\kappa_r$ and $\kappa_\theta$ denote the radial and azimuthal curvatures of the undeformed shape respectively. In the case of a spherical shell, where both curvatures of the underformed shape are constant, the terms in the round parenthesis in the last row are constant and the integral simplifies to null (once the simplifications listed above are used). Using the assumption of a narrow ridge ($s+r_d\approx r_d$), it can also be shown that the integral simplifies to null for a conical shape since $z_0''=0$ and $z_0'=const$. In general, we can assume that this integral simplifies to null for all shapes of revolution where $z_0''$ and $z_0'$ do not vary significantly within the ridge region and we can approximate the undeformed curvatures in the ridge region with their values at the ridge location, thus reducing the terms in the round parenthesis to constant. In this scenario, the energy in Eqn~\eqref{EQ:EnergyKN} simplifies to the energy presented in \cite{Knoche2014} and the ridge is symmetric about $s=0$, because the variational problems to be solved in the regions $[-\xi_i,0]$ and $[0,\xi_o]$ are the same. 

We introduce the nondimensionalization:
\begin{equation}\begin{split}\label{eq:nondim_pogorelov}
s = \zeta \bar{s};\quad u = \zeta \beta^2 \bar{u}; \quad v'(s) = \beta \bar{w};\\ \quad \zeta  =\left(\frac{t^2}{12}\right)^{1/4}r_d^{1/2}\beta^{-1/2},
\end{split}\end{equation}
where the parameter $\beta = |z_0'(r_d)|$ describes the local slope of the undeformed shape at the ridge position, $r_d$. Using this nondimensionalisation the problem recasts in the same dimensionless form as \cite{Knoche2014} and can be minimized to obtain the same dimensionless solution presented therein (and not reported again here). In dimensionless form the functions $\bar{u}(\bar{s})$ and $\bar{w}(\bar{s})$ which describe the shape of the ridge, are thus valid for any shell of revolution with non-negative Gaussian curvature and  $z_0''$ and $z_0'$ which do not vary significantly within the ridge region.

\subsection{Mechanics of the ridge} \label{sec:localSlope}

Following the approach of \cite{Knoche2014}, we estimate the geometrical and mechanical properties of the ridge. In the inner region of the ridge we approximate $K_r(r) \approx a_c (r-r_c)$, and the hoop stress  $\tau_{\theta} \approx -\tau_0 [1-a_p (r-r_c)^2]$, where $r=r_c$ is the radial coordinate where the radial curvature vanishes. This approximation is useful because the parameters $a_p$, $a_c$, and $\tau_0$ have precise geometrical and mechanical meanings and can be computed from the analytical solution for the ridge described in Section~\ref{sec:ridgeStructure}. The parameter $a_p$ is the inverse square of the ridge width, $a_c$ is related to the ridge curvature, and $\tau_0$ is the maximum compressive hoop stress in the ridge. To compute these quantities the following conditions are imposed (more details can be obtained from \cite{Knoche2014}): (i) the minimum approximated hoop stress is equal to the minimum analytical hoop stress, which fixes $\tau_0$; (ii) the integral of the hoop stress in the negative (compressive) region is the same in both the approximated and analytical cases, which fixes $a_p$; and (iii) recalling that the meridional curvature in the inner region is $K_r (r)= - 2 z_0''(r) +v''(r)$, the coordinate $r_c$ is found by imposing $K_r (r_c)=0$, so that $a_c = \upd K_r (r)/\upd r \vert_{r=r_c}$.

With this analytical framework we can predict how the hoop stress, $\tau_0$, and geometry of the ridge (i.e. the width $w_r$ and the characteristic radius of curvature $r_{curv}$) scale in terms of the local slope of the ridge, $\beta$, and the radial position of the ridge, $r_d$. For any shell of revolution with non-negative Gaussian curvature, we have
\begin{equation}\label{EQ:scalings}
\begin{split}
\tau_{0} &\approx c_{\tau}E \frac{t^{3/2}\beta^{3/2}}{\sqrt{r_d}} \quad w_r \approx \frac{2}{c_p^{1/2}12^{1/4}}\frac{\sqrt{t\,r_d}}{\beta^{1/2}} \\
r_{curv} &\approx \frac{c_p^{1/2}}{c_c 12^{1/4}}\frac{\sqrt{t\, r_d}}{\beta^{3/2}}.
\end{split}
\end{equation}
It can be shown that the constants $c_p = 0.33955$, $c_{\tau} = \sigma \left(12^{-1/4}\right)\left[3(1-\nu^2)\right]^{-1}$, and $\sigma =  1.2466$  are valid for all geometries \cite{Knoche2014}, while, in general, the constant $c_c$ must be computed numerically for each shape, except for a cone where $c_c =  \sigma\left(e^{-\frac{\pi }{4}}\right)/3$.

We let $\delta$ be the displacement of the shell apex in the negative $z$ direction, i.e. the indentation depth. For any mirror buckled shallow shell of revolution we can write $\delta = c_r \beta r_d$, with $c_r = 1$ for a cone and $c_r=1/2$ for a sphere. From geometrical considerations, for the other shells investigated in this work we expect that $c_r=c_r(r_d)$ is not constant but always bounded between $1/2$ and $1$. This parameter will be important when we compare the behaviour of different shapes in Section \ref{sec:circumBuckling}.

In the case of a shallow mirror buckled cone, $\beta = const$, $r_d \sim \delta/\beta$, and the ridge has the following properties:
\begin{align}\label{EQ:scalingsdeltaCone}
	\tau_0 \sim E \frac{t^{3/2}\beta^2 }{\sqrt{\delta}} &&  w_r \sim \frac{\sqrt{t \delta}}{\beta} && r_{curv} \sim \beta^{2}\sqrt{t \delta},
\end{align}
which is consistent with the results of Seffen \cite{Seffen2016a}. In the process of deriving the energy scaling of a circular ridge, Witten \cite{Witten2007} proposed that a ridge formed by partially inverting a cone accommodates a radius of curvature $r_{curv}\sim\sqrt{t r_d}$ which matches with our scaling in \eqref{EQ:scalingsdeltaCone}, since $\beta$ is constant. However, the scaling in \cite{Witten2007} cannot be readily generalized to any shape of revolution because $\beta$ depends on the radial location of the ridge (or, equivalently on $\delta)$. Interestingly, in the case of a cone increasing $\delta$ causes the compressive hoop stress to decrease in magnitude. The ridge also becomes wider and less curved. 

For a mirror buckled shallow spherical shell of radius $R$, $\beta \sim r_d/R$ and $r_d \sim \sqrt{\delta R}$ which gives
\begin{align}\label{EQ:scalingsdeltaSphere}
	\tau_0 \sim E \frac{t^{3/2} \delta^{1/2}}{R} && w_r \sim \sqrt{t\, R} && r_{curv} \sim t\, \delta^{-1/4}R^{5/4},
\end{align}
consistently recovering \cite{Knoche2014}. In contrast to the cone, when indenting a sphere the hoop stress increases, the ridge maintains a constant width and also becomes less curved.  This is important for the circumferential buckling of the ridge and will be discussed in Section~\ref{sec:results}.

To derive the strain energy of the ridge, we write the circumferential strain, $\gamma$, as
\begin{equation}\label{eq:strain}
\gamma \approx \frac{\tau_0}{Et} \sim  \frac{t^{1/2}\beta^{3/2}}{r_d^{1/2}} \approx c_r^{-1/2}\beta^2\sqrt{\frac{t}{\delta}},
\end{equation}
and the characteristic radius of curvature as 
\begin{equation}\label{eq:radiuscurvature}
r_{curv} \approx \frac{\sqrt{t r_d}}{\beta^{3/2}} \sim c_r^{1/2}\frac{\sqrt{t \delta}}{\beta^{2}}
\end{equation}
which both depend on the local slope at the ridge location, the indentation depth, and the function $c_r(r_d)$ as defined earlier. The total energy of a circular ridge in a mirror buckled shell of revolution thus scales with the radial position and the local slope, or the apex displacement and the local slope, as 
\begin{equation}\label{eq:energy}\begin{split}
E_n &\approx r_d E t w_r \gamma^2+ E t r_d w_r \left(\frac{t}{r_{curv}}\right)^2\\ &\sim E t^{5/2}r_d^{1/2}\beta^{5/2} \sim E t^{5/2}c_r^{-1/2}\delta^{1/2}\beta^2.
\end{split}\end{equation}
where in the last step we used the relation $\delta = c_r \beta r_d$ proposed earlier in this section. By choosing the local slope, $\beta$, that corresponds to the specific shape, this energetic form recovers both \cite{Seffen2016a} and \cite{Witten2007} for a cone and \cite{Knoche2014} for a depressurised sphere (deforming axisymmetrically via mirror buckling).
\section{Shallow shells of revolution}\label{sec:shallowShell}

In Section~\ref{SEC:PogorelovCone} we derived the geometrical and mechanical properties of the circular ridge forming in mirror buckled shells of revolution without discussing if (and how) a shell accommodates a mirror buckled shape. Here we consider the indentation problem, where the apex of a shallow shell of revolution (with non-negative Gaussian curvature) is displaced continuously in the negative $z$ direction, and evaluate the onset of circumferential buckling of the ridge, without assuming a mirror buckled form a priori.

We consider the same cylindrical coordinate system introduced in Section~\ref{SEC:PogorelovCone}, and the same undeformed mid-profile, $z_0(r)$, for a general shell of revolution. We call $w(r,\theta)$ the displacement in the $z$ direction (out-of-plane) and $\phi(r,\theta)$ the Airy stress function. For a shallow shell the equilibrium and compatibility equations read \cite{Ventsel2001}:
\begin{subequations}\label{EQ:ShallowShells}
\begin{align}
		B\nabla^4 w + \nabla_k^2 \phi - \Lop\left(\phi,w\right) &= 0, \label{EQ:force_balance}\\
		\frac{1}{Et}\nabla^4 \phi + \tfrac{1}{2} \Lop\left(w,w\right) - \nabla_k^2 w &=0.\label{EQ:compatibility}
\end{align}
\end{subequations}
where $\nabla^2\left(\cdot\right)$ is the Laplacian operator in polar coordinates and the operator $\Lop(\cdot,\cdot)$ is defined as
\begin{equation}\label{EQ:OperatorN}
\begin{split}
\Lop\left(f,g\right)=\frac{\partial^2 f}{\partial r^2}\left(\frac{1}{r}\frac{\partial g}{\partial r} + \frac{1}{r^2} \frac{\partial^2 g}{\partial \theta^2}\right)+\frac{\partial^2 g}{\partial r^2}\left(\frac{1}{r}\frac{\partial f}{\partial r} + \frac{1}{r^2} \frac{\partial^2 f}{\partial \theta^2}\right) \\- 2 \frac{\partial}{\partial r}\left(\frac{1}{r} \frac{\partial f}{\partial \theta}\right) \frac{\partial}{\partial r}\left(\frac{1}{r} \frac{\partial g}{\partial \theta}\right).
\end{split}
\end{equation}
The Vlasov operator for a shallow shell of revolution rewrites as
\begin{equation}\begin{split}\label{eq:VlasovOperator}
\nabla_k^2 (\cdot) = \kappa_{\theta}\frac{\partial^2 \left(\cdot\right)}{\partial r^2} + \kappa_r \left(\frac{1}{r}\frac{\partial (\cdot)}{\partial r} + \frac{1}{r^2}\frac{\partial^2 (\cdot)}{\partial \theta^2}\right).
\end{split}\end{equation}
We define $\cH$ and $\cL$ as the characteristic lengths in the out-of-plane and in-plane directions. Then, after introducing the following nondimensionalization,
\begin{align}\label{EQ:nondimensionalization_general}
	W = \frac{w}{\cH} ; &&  \rho =\frac{r}{\cL}; && \Phi = \frac{\phi}{E t \cH^2}; && \left(\bar{\kappa}_r,\bar{\kappa}_{\theta}\right)=\frac{\cL^2}{\cH} \left(\kappa_r,\kappa_{\theta}\right)
\end{align}
the shallow shell equations \eqref{EQ:ShallowShells} rewrite as
\begin{subequations}\label{EQ:ShallowShells_normalizedFULL}
	\begin{align}
		\frac{1}{\lambda^4}\nabla^4 W + \bar{\nabla}_k^2 \Phi - \Lop \left(\Phi, W\right) &=0 \label{EQ:vertical_force_normalizedFULL}\\
		\nabla^4 \Phi + \tfrac{1}{2} \Lop\left(W,W\right) - \bar{\nabla}_k^2 W&=0\label{EQ:compatibility_normalizedFULL}
	\end{align}
\end{subequations}
where the parameter $\lambda$ is defined as
\begin{equation}\label{eq:lambdaC}
\lambda = \left(12\left(1-\nu^2\right)\right)^{1/4}\sqrt{\frac{\cH}{t}}.
\end{equation} 
This parameter is related to the ratio between the stretching and bending energies involved in achieving a mirror buckled shape (see \cite{Taffetani2018,Lobkovsky1996a} for a discussion in the case of a spherical cap) \rp{with $\lambda^4$ being the F\"oppl von K\'arm\'an number \cite{Knoche2014}}. The $\Lop$ and $\nabla^4$ operators are defined as before with the dimensional coordinates replaced by their dimensionless counterparts. The Vlasov operator rewrites in dimensionless form as
\begin{equation}
\bar{\nabla}_k =  \bar{\kappa}_{\theta}\frac{\partial^2 \left(\cdot\right)}{\partial \rho^2} + \bar{\kappa}_r \left(\frac{1}{\rho}\frac{\partial (\cdot)}{\partial \rho} + \frac{1}{\rho^2}\frac{\partial^2 (\cdot)}{\partial \theta^2}\right).
\end{equation}

\subsection{Circumferential Buckling of an Axisymmetric Ridge}\label{sec:circumBuckling}

The quantitative analysis of circumferential buckling is performed using a linear stability analysis on the shallow shell equations \eqref{EQ:ShallowShells} following standard steps (see \cite{Taffetani2018} for example). We first write the solution as an axisymmetric contribution over which we superimpose a small ($\epsilon \ll 1$) sinusoidal perturbation in the circumferential direction as 
\begin{equation}\begin{split}\label{EQ:ansatzLSA}
W = W^{(0)}(\rho) + \epsilon W^{(1)}(\rho) \cos  m \theta;\\ \Phi = \Phi^{(0)}(\rho) + \epsilon \Phi^{(1)}(\rho) \cos m \theta.
\end{split}\end{equation}
We impose symmetry conditions at the apex:
\begin{equation}\label{eq:symmetry_bcs}
\begin{split}
W|_{\rho=0} &=\Delta_0; \quad \left.\frac{\partial W}{\partial \rho}\right\vert_{\rho=0}=0;\\
\left.U_r\right\vert_{\rho=0}&=0; \quad \left.\Phi\right\vert_{\rho=0} = 0,
\end{split}
\end{equation}
with $\Delta_0$ being the dimensionless applied displacement of the apex and $U_r$ denoting the dimensionless radial displacement. Clamped conditions are applied to the outer boundary as 
\begin{equation}
\begin{split}
\left.W\right\vert_{\rho=1} &=0; \quad \left.\frac{\partial W}{\partial \rho}\right\vert_{\rho=1}=0;\\
\left.U_r\right\vert_{\rho=1}&=0; \quad \left.U_{\theta}\right\vert_{\rho=1} = 0
\end{split}
\end{equation}
with $U_{\theta}$ denoting the dimensionless circumferential displacement. The boundary conditions are applied to both the axisymmetric and $O(\epsilon)$ problems, as described in \cite{Taffetani2018} for the spherical case. The condition $U_r|_{\rho=0}=0$ for the axisymmetric problem in a generic shell of revolution can be rewritten as 
\begin{equation}
\lim_{\rho\rightarrow 0} \left[\rho\frac{\upd^2 \Phi}{\upd \rho^2} - \nu \frac{\upd \Phi}{\upd \rho} - \Delta_0 |Z_0(\rho)|\right]=0
\end{equation} 
where the last term reduces to zero for the spherical case.

We solve the leading order axisymmetric problem to find the functions $W^{(0)}(\rho)$ and $\Phi^{(0)}(\rho)$ using the function \emph{bvp4c} in Matlab. Finally, we recast the problem at order $O(\epsilon)$ as a second order polynomial eigenvalue problem to compute, using the function \emph{polyeig} in Matlab, the nondimensional critical indentation depth, $\Delta_{c} = \delta_{c}/\cH$, when the ridge undergoes circumferential buckling into a polygonal shape with $m_{c}$ sides. These results are compared with the expected behaviour obtained from a near threshold analysis, assuming a mirror buckled form, described in the next section.

\subsection{Buckling of the Ridge in a Mirror Buckled Shallow Shell of Revolution}\label{sec:bucklingOfPRidge}

Equation \ref{EQ:ShallowShells_normalizedFULL}(a) can be rewritten in terms of the dimensionless stresses, $\Sigma_{rr}$, $\Sigma_{\theta\theta}$, and $\Sigma_{r \theta}$ as
\begin{equation}\begin{split}\label{EQ:Shallow_Force_Stress}
\frac{1}{\lambda^4}\nabla^4 W + \Sigma_{\theta\theta}\left(\bar{\kappa}_{\theta} -\frac{1}{\rho}\frac{\partial W}{\partial \rho}- \frac{1}{\rho^2}\frac{\partial^2 W}{\partial \theta^2}\right) \\+ \Sigma_{\rho\rho}\left(\bar{\kappa}_{r} -\frac{\partial^2 W}{\partial \rho^2}\right) + 2\Sigma_{\rho\theta}\frac{\partial }{\partial \rho}\left(\frac{1}{\rho}\frac{\partial W}{\partial \theta}\right)  =0.
\end{split}\end{equation}
Inserting the small perturbation ansatz, \eqref{EQ:ansatzLSA}, into \eqref{EQ:Shallow_Force_Stress} and looking at the first order expansion of Eqn \eqref{EQ:Shallow_Force_Stress}, we can use near threshold scaling arguments to predict the circumferential buckling behaviour of the ridge in a \textit{mirror buckled} shallow shell of revolution.

In Section~\ref{SEC:PogorelovCone} we fully characterised the circular ridge of a mirror buckled shallow shell of revolution. From Eqns~\eqref{EQ:scalings} and \eqref{eq:strain}, denoting $\rho_d$ as the dimensionless radial position of the ridge, the dimensionless ridge width, $W_r$, and strain, $\Gamma$, rewrite as
\begin{equation}\begin{split}
W_r \sim \lambda^{-1} \rho_d^{1/2} \hat{Z}^{-1/2} \\ \Gamma \sim \lambda^{-1}  \rho_d^{-1/2}\hat{Z}^2,
\end{split}\end{equation}
where we used the rescaling in Eqns~\eqref{EQ:nondimensionalization_general} and the local slope at the ridge location is rescaled as $\beta = \left(\mathcal{H}/\mathcal{L}\right) \hat{Z}$. For compactness, $\hat{Z} = Z'(\rho_d)$ hereafter. 

The only non-zero strain in the ridge is the hoop strain, $\Gamma$, and thus $\Sigma_{rr},\Sigma_{\theta \theta}\sim \Gamma$ and $\Sigma_{r\theta}=0$. The length scale over which the first order quantities vary is $W_r$ and the characteristic radial coordinate is the position of the ridge, $ \rho_d$. We can identify the three dominant terms: (i) bending in the circumferential direction, $\sim m^4 \lambda^{-4} \rho_d^{-4} W^{(1)}$; (ii) the (compressive) hoop stress, $\sim \Gamma m^2 \rho_d^{-2} W^{(1)}\sim m^2 \lambda^{-1} \rho_d^{-5/2} \hat{Z}^2   W^{(1)}$; and (iii) the radial stress, $\sim \Gamma W_r^{-2} W^{(1)} \sim \lambda \rho_d^{-3/2} \hat{Z}^{3} W^{(1)}$. Balancing these terms, we expect the ridge of a mirror buckled shallow shell of revolution undergoes circumferential buckling with a mode number $m_c \sim \lambda \rho_d^{1/2} \hat{Z}^{1/2}$ at a dimensionless critical radius $\rho_d \sim \lambda^{-2} \hat{Z}^{-1}$. So we expect the critical mode $m_c\sim const$. As a consequence the following scalings hold at the onset of buckling assuming a mirror buckled form:
\begin{equation}\label{EQ:scalingsbuckling}\begin{split}
T_0 \sim \hat{Z}^2, \quad \rho_{curv}^{-1} \sim \lambda^2 \hat{Z}^2, \quad \Delta_c \sim \lambda^{-2}, \\ W_r \sim \lambda^{-2}\hat{Z}^{-1}, \quad \rho_d \sim \lambda^{-2} \hat{Z}^{-1}, \quad \quad m_c\sim const.
\end{split}\end{equation}
where $T_0$ and $\rho_{curv}$ are the dimensionless characteristic hoop stress and dimensionless radius of curvature, respectively.

\section{Circumferential buckling of shallow shells of revolution with constant Gaussian curvature}\label{SEC:circumBuckling} 

We focus on the particular subset of shallow shells of revolution that have constant Gaussian curvature such that 
\begin{equation}\label{eq:kG_shallow_dimensional}
\kappa_G = \kappa_{\theta}\kappa_r = \frac{1}{r}\frac{\upd z_0}{\upd r}\frac{\upd^2 z_0}{\upd r^2} = const.
\end{equation}
If we use the rescaling in \eqref{EQ:nondimensionalization_general}, we can rewrite the differential equation in \eqref{eq:kG_shallow_dimensional} as
\begin{equation}\label{eq:kG_shallow}
\frac{1}{\rho}\frac{\upd Z_0}{\upd \rho}\frac{\upd^2 Z_0}{\upd \rho^2} = \bar{\kappa}_G
\end{equation}
with $Z_0(\rho)$ being the dimensionless undeformed mid-profile of the shell.

By a suitable choice of the dimensionless Gaussian curvature, $\bar{\kappa}_G$, we can recover the familiar limiting cases of a conical shell ($\bar{\kappa}_G=0$) and a spherical cap with radius $R$ ($\bar{\kappa}_G$=1). In between these two values we obtain a series of shapes bounded from above by the reference spherical shell and from below by the conical shell inscribed into the spherical one (i.e. its mid-profile is the chord that connects the apex to the outer boundary of the spherical shell). The mid-profiles, obtained by solving Eqn~\eqref{eq:kG_shallow} while fixing the base to be $Z(1)=0$, \rp{are given by
	\begin{equation}\begin{split}
	Z_0(\rho) = &\frac{1}{2} \left(\sqrt{\bar{\kappa}_G + Z'(0)^2}-\rho \sqrt{\bar{\kappa}_G \rho^2+Z'(0)^2}\right) - \\
	& \frac{Z'(0)^2}{2\sqrt{\bar{\kappa}_G}}\log \left(\frac{\bar{\kappa}_G \rho + \sqrt{\bar{\kappa}_G} \sqrt{\bar{\kappa}_G \rho^2+Z'(0)^2}}{\bar{\kappa}_G + \sqrt{\bar{\kappa}_G} \sqrt{\bar{\kappa}_G + Z'(0)^2}}\right)
	\end{split}\end{equation}
	Using the constraint on the overall shell rise, $Z(0)=0.5$, the slope at the apex can be obtained numerically as $Z'(0) = [-0.5, -0.41, -0.32, -0.21, 0.0]$ for $\bar{\kappa}_G = [0, 0.25, 0.5, 0.75, 1]$ respectively.} These shapes are sketched in Fig~ \ref{fig:shapes}(a). With these geometrical choices, we consider shells with the same rise $\cH$ and the same planform radius $\cL$. A feature of these shapes, seen by looking at the function $Z'(\rho)$ --- shown in Fig~ \ref{fig:shapes}(b), is that all intermediate shapes transition from a cone-like shape at the apex, where $Z''(\rho=0)=0$, to a sphere-like shape at the outer boundary.

We consider these mid-profiles to evaluate shells subjected to indentation of the apex for two reasons: to evaluate the mirror buckling hypothesis for general shells of revolution at the circumferential buckling threshold, and to investigate the properties of ridges at radial coordinates with different local slopes. 

\subsection{Comparing the Buckling of a Ridge with Different Mid-profiles}\label{sec:GeneralbucklingOfPRidge}
\begin{figure}[h!]
	\centering
	\includegraphics[width=0.9\textwidth]{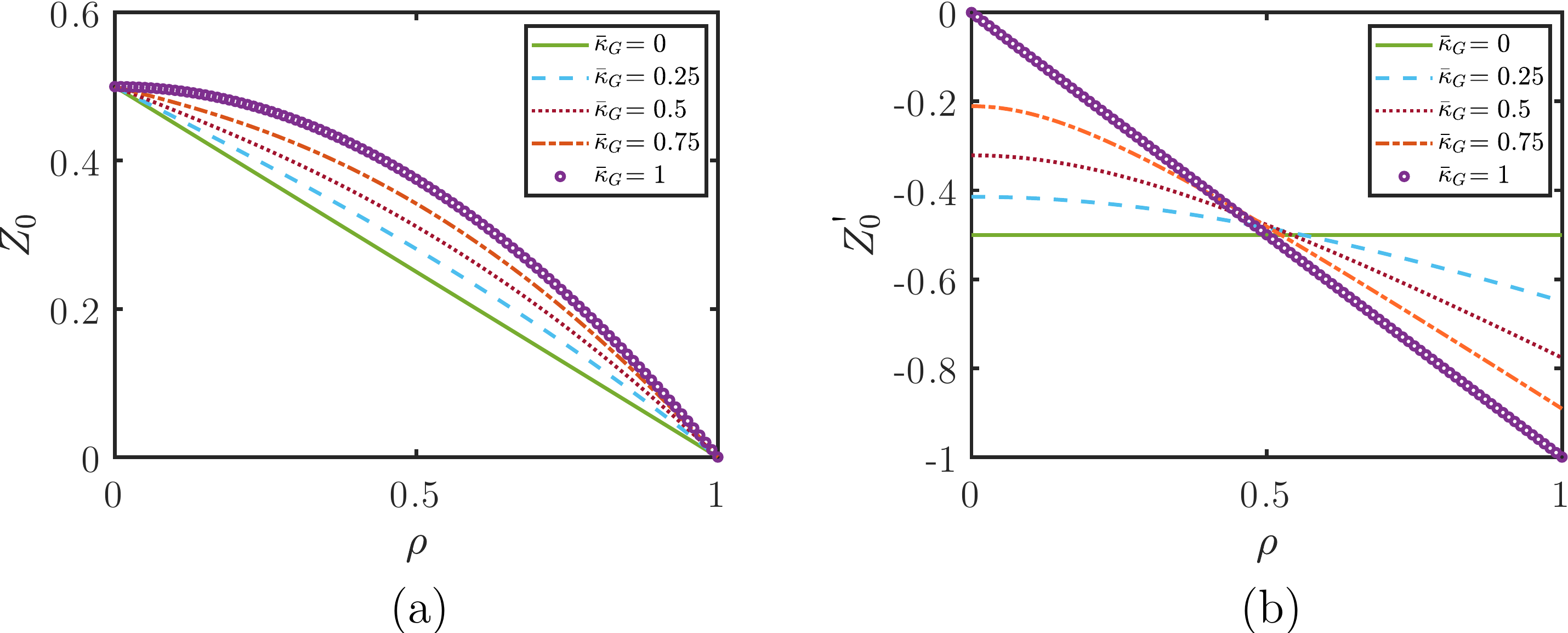}
	\caption{ (a) Mid-profile $Z_0(\rho)$ and (b) first derivative of the mid-profiles, $Z'_0(\rho)$, of the five shells of revolution with constant Gaussian curvature. The shapes are bounded from above by the reference spherical shell ($\bar{\kappa}_G=1$) and from below by a conical shell ($\bar{\kappa}_G=0$) inscribed into the spherical one. A feature of these shapes, shown in (b), is all intermediate shapes transition from a cone-like shape at the apex\rp{, where the second derivative of the mid-profile is null --- characteristic of a conical profile everywhere,} to a sphere-like shape at the outer boundary\rp{, where they approach a value of $-1$ --- typical of the dimensionless spherical profile}.}   \label{fig:shapes} 
\end{figure}

The unified description for the features of the axisymmetric ridge across different shells of revolution relies on the local slope of the shell at the ridge location, effectively reducing each shell at each deformed stage to an equivalent conical shell with the same slope throughout the entire radial span. The natural way to rescale all such cones, with different angles $\beta$, is to take $\cL=L$ and $\cH=L\beta$ in Section \ref{sec:shallowShell}; with this choice the scalings proposed in Eqns~\eqref{EQ:scalingsbuckling} would not depend on any additional geometric factors. However, the radial location where the ridge buckles is not known \emph{a priori}, so we cannot use $\beta$ to rescale all the mid-profiles used in this analysis. This is the reason why a non-dimensionalization based on the rise and the size of the planform is used, since they are fixed. However, to obtain a consistent comparison across the mid-profiles shown in Fig.~\ref{fig:shapes}(a), the scalings in Eqns~\eqref{EQ:scalingsbuckling} must be updated to include the geometric factor $c_r$, which is computed as
\begin{equation}
c_r=c_r(\rho_d) = \frac{Z(0)-Z(\rho_d)}{\rho_d \hat{Z}}.
\end{equation}
This term was already introduced in Section~\ref{sec:localSlope}, being the same parameter that relates the radial position of the ridge to the indentation depth once the local slope is known; it can also be interpreted as how much \emph{smaller} the rise of the profile $Z(\rho_d)$ is compared to the equivalent conical shell constructed using the local slope at $\rho = \rho_d$. Naturally, for a conical mid-profile, $c_r=1$.

We can define $\cH_{effective}$ as the rise of the equivalent conical mid-profile, so that $c_r \cH_{effective}$ is the actual rise of the mid-profile under analysis. Using the definition of $\lambda$ in Eqn~\eqref{eq:lambdaC}, and employing this new rise, we can write  $\lambda_{effective} = \lambda c_r^{-1/2}$ and $\Delta_{c,effective} = \Delta_{c} c_r$. Incorporating the geometrical factor, $c_r$, the scalings in Eqn~\eqref{EQ:scalingsbuckling} thus become
\begin{equation}\begin{split}\label{eq:buckling_scalings}
T_0 \sim c_r^{-1/2}\hat{Z}^2, \quad \rho_{curv}^{-1} \sim  c_r^{1/2}\lambda^2 \hat{Z}^2, \\
\Delta_c \sim c_r^2 \lambda^{-2}, \quad W_r \sim c_r\lambda^{-2}\hat{Z}^{-1}, \\ \rho_d \sim c_r\lambda^{-2} \hat{Z}^{-1}, \quad m_c\sim const.
\end{split}\end{equation}

\subsection{Results and Discussion}\label{sec:results}
\rp{\subsubsection{Ridge behaviour at the buckling threshold}}
We expect buckling of the ridge to occur when the compressive hoop stress in the inner edge of the ridge exceeds the buckling capacity of that region. However, as discussed in Section~\ref{sec:localSlope}, the hoop stress, ridge width, and curvature change with both the ridge position and local slope, but in different ways for different mid-profiles. From the buckling of cylindrical shells under axial load, we expect increasing the ridge radius of curvature to lead to a lower critical buckling stress \cite{Calladine1983}. In spherical shells the hoop stress increases monotonically with indentation while the ridge radius of curvature decreases, see Eqns \eqref{EQ:scalingsdeltaSphere}. In contrast, for a conical shell the hoop stress decreases with indentation, the ridge radius of curvature increases, see Eqns \eqref{EQ:scalingsdeltaCone}. Buckling of the ridge is therefore due to the interplay between the changing hoop stress and geometry of the ridge.

From Eqn~\eqref{eq:energy}, the strain energy of the ridge scales with the local slope, $\beta$, and length of the ridge,  $l = 2 \pi r_d$, as $E_n \sim \beta^{5/2} l^{1/2}$. It has previously been proposed that the increasing angle of the ridge as a spherical shell is indented increases the ridge strain energy thus leading to buckling when the straight-ridge form becomes more energetically favourable \cite{Nasto2013, Lobkovsky1996a}; indeed the energy of a straight ridge is $E_{n, straight} \sim \beta^{7/3} l^{1/3}$ \cite{Witten2007,Lobkovsky1996a}. A deeper inspection of the energy scaling, however, reveals that there are two mechanisms which can potentially control circumferential buckling of a ridge into a polygonal shape: the local slope and the length of the ridge. This latter quantity explains why the circular ridge in a conical geometry eventually buckles even though the slope remains constant. This observation is consistent with our analytical results which show how the quantities important for buckling, in particular the compressive hoop stress and ridge curvature, scale with both the local slope and ridge radius. We also show how the geometry of the shell, through the parameter $\lambda$, influences buckling. 

To evaluate the validity of our model, and the mirror buckling assumption, we compare the predicted ridge properties at the threshold of circumferential buckling to those obtained by a linear stability analysis on the shallow shell equations (Section~\ref{sec:shallowShell}). The analysis is carried out on the five, constant Gaussian curvature, mid-profiles presented in Fig.~\ref{fig:shapes} with the parameter $\lambda$ spanning the range $[17.5, 100]$. The lower limit is chosen to avoid scenarios where the ridge buckles close to the outer boundary and the upper limit is chosen for numerical stability reasons, discussed later in this Section. \rp{These shells are shallow and very thin. Physical examples at the nanoscale, where gravity has a weaker effect and thin objects can self sustain, include polymeric vesicles (polymersomes) - of interest for drug delivery and diagnoistic applications \cite{Matoori2020,Wong2019,LeMeins2011}, and non-biological hollow nanoshells \cite{Yu2018} both with radii between $50-100$~nm and thicknesses between $3-5$~nm. While these are generally closed and deep shells, under suitably shallow indentations our analysis, based on shallow and clamped shells, would be applicable.}

\begin{figure}
	\centering
	\includegraphics[width=0.85\textwidth]{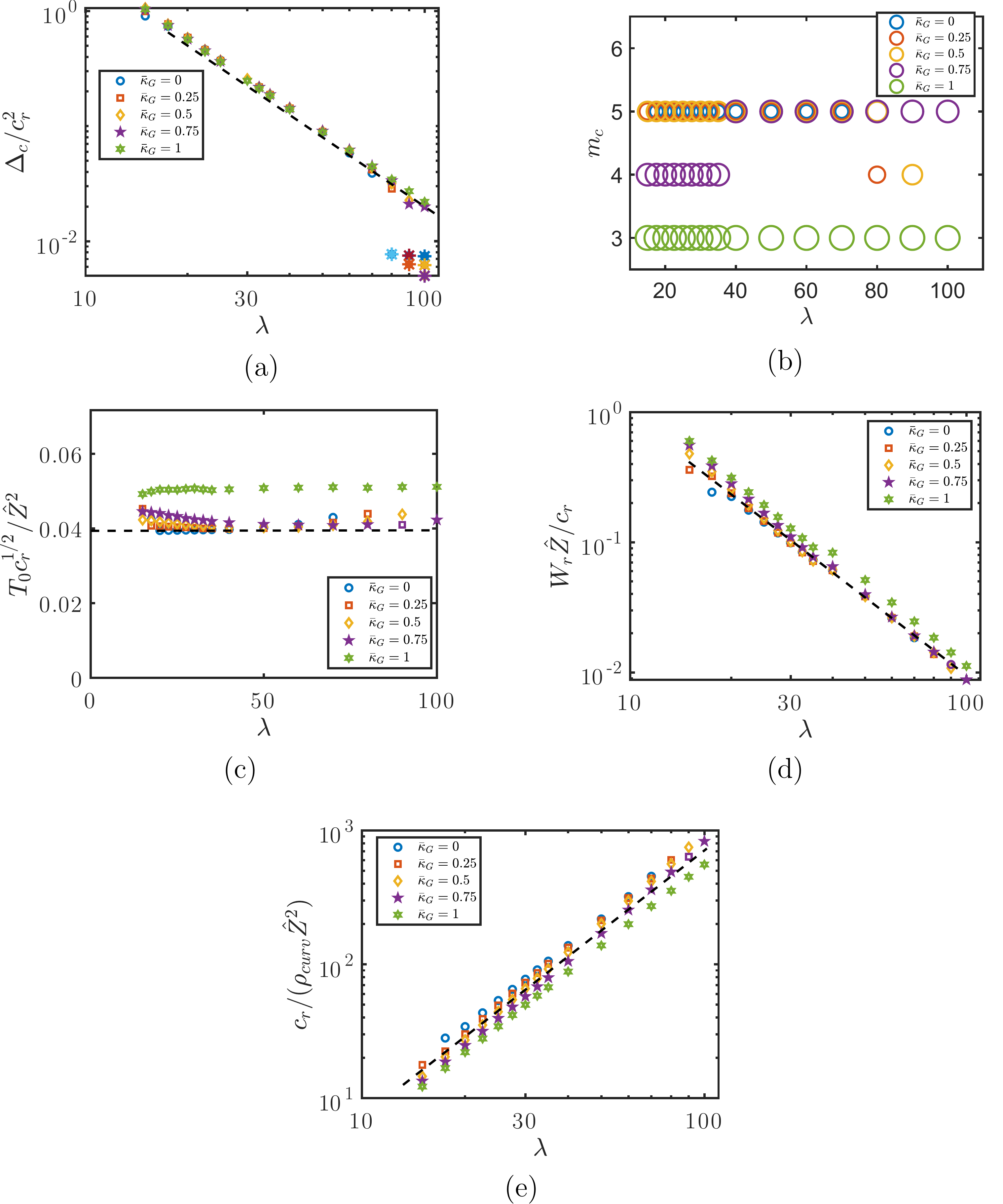}
	\caption{Ridge properties at the onset of circumferental buckling for shells with  constant Gaussian curvature rescaled using Eqns \eqref{eq:buckling_scalings}.  The markers show the linear stability analysis results and the dashed lines indicate the expected behaviour against $\lambda$ assuming mirror buckling. The critical indentation  (a) shows excellent agreement with the expect behaviour. The critical circumferential buckling mode, $m_c$, is shown in (b). The maximum compressive hoop stress (c); width of the ridge (d); and ridge curvature (e) all show very good agreement with the expected behaviour obtained assuming a mirror buckled form.}   \label{fig:K_effect_scaled}
\end{figure}

The raw data for the ridge properties at the onset of circumferential buckling are extracted from the numerical solution of the shallow shell equations and are included in Appendix \ref{AppendixA}. The data are rescaled using the relations in Eqns~\eqref{eq:buckling_scalings} and shown in Fig.~\ref{fig:K_effect_scaled}. When including the effect of the local slope, via  $\hat{Z}$, and the geometrical factor $c_r$, the geometries analysed collapse onto the unified behaviour predicted from our theoretical framework assuming a `mirror buckled' shape (shown as dashed lines). This suggests that the `mirror buckling' assumption  describes the axisymmetric deformation of these shells well at the onset of circumferential buckling except, partially, the spherical case ($\kappa_G = 0$). \rp{It is important to highlight that the quantities $\hat{Z}$ and $c_r$ used to non-dimensionalise the plots in Fig.~\ref{fig:K_effect_scaled} are the values of the actual ridge obtained from the numerical solution of the shallow shell equation rather than those obtained using the mirror buckling assumption once the indentation depth is prescribed. We will evaluate the limitations of the mirror buckling assumption in the next section, but the results shown in Fig.~\ref{fig:K_effect_scaled} reveal how the properties of the ridge are controlled by the local slope at the ridge location.}

Figure \ref{fig:K_effect_scaled}(a) shows the rescaled critical indentation as a function of $\lambda$ and follows the predicted behaviour from our analytical model extremely well. The linear stability results were validated against finite element analysis showing very good agreement across the full range of parameters considered, see Section~\ref{sec:FEA}. The data denoted by an asterisk in \ref{fig:K_effect_scaled}(a) are not reliable due to numerical instabilities ---  in these cases the numerical solution of the eigenvalue problem did not provide a unique stability threshold. For clarity we have removed the corresponding data from the remaining plots. 
Figure~\ref{fig:K_effect_scaled}(d) shows the width of the ridge, which is computed as the region where the hoop stress in negative, i.e. half the ridge width since only the inner region is under compression. The ridge width decreases with increasing $\lambda$ and matches the predicted behaviour very well. Figure~\ref{fig:K_effect_scaled}(e) shows the ridge curvature (computed at the radial position $\rho_d$). The curvature of the ridge increases with increasing $\lambda$. These results show that, with the appropriate rescaling, predictions made using the mirror buckling assumption match the behaviour of our shells of revolution very well. 

The critical circumferential buckling mode of the ridge, shown in Fig.~\ref{fig:K_effect_scaled}(b), and the maximum (compressive) hoop stress in the ridge region, shown in Fig~\ref{fig:K_effect_scaled}(c), both show some deviation from the prediction. Our analysis in Section~\ref{sec:bucklingOfPRidge} shows that the ridge always buckles with a constant number of straight ridges independent of the geometry of the shell and the local slope (embedded, in our description, via $\lambda$ and $\hat{Z}$). In the case of the deflation of spheres, Kierfield and Knoche \cite{Knoche2014}, show that circumferential buckling of mirror buckled deflated spherical shells usually occurs with a mode number of $5$ across a wide range of reduced stiffness (equivalent to our $\lambda$). We obtain the same mode number (5) by indenting shells with constant Gaussian curvature $\bar{\kappa}_G =0, 0.25, 0.5$. However, for $\bar{\kappa}_G =0.75$ a transition from $m_c=5$ to $m_c=4$ is observed, and for the spherical case a mode number $m_{c}=3$ emerges (confirmed by experiments in \cite{Nasto2013}).  The maximum (compressive) hoop stress in the ridge region is shown in Fig~\ref{fig:K_effect_scaled}(c). This quantity does not collapse as well as the other properties, especially for the spherical case, although the constant behaviour with $\lambda$ predicted from the near threshold analysis is evident in all the cases after the effect of both the local slope, $\hat{Z}$, and the geometrical factor, $c_r$, are included.

\begin{figure*}[!ht]
	\centering
	\includegraphics[width=0.9\textwidth]{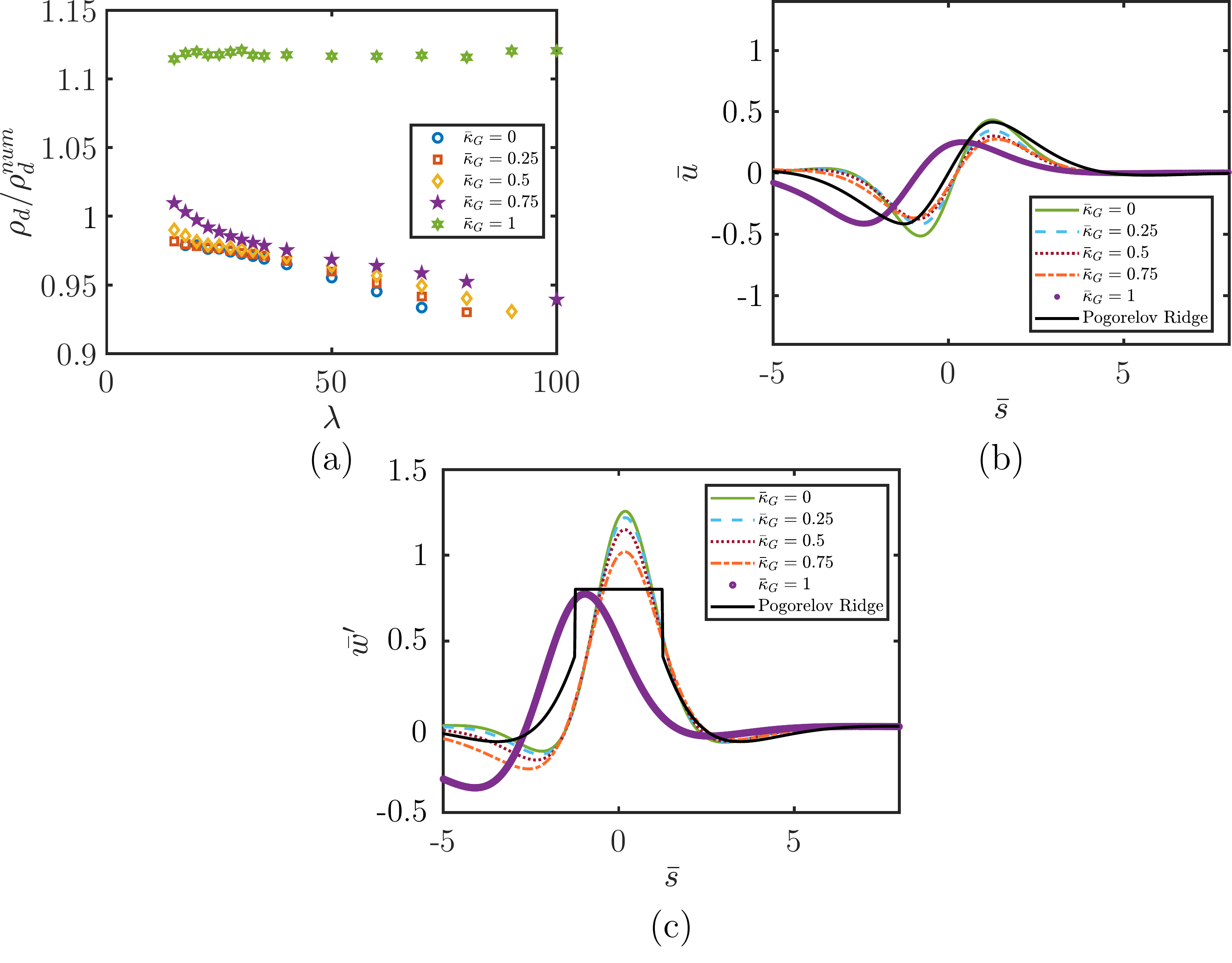}
	\caption{Comparison of the ridge between the analytical solution from the Pogorelov approach (Section \ref{sec:localSlope}) and the solution obtained from the axisymmetric shallow shell equations in terms of the (a) ridge radius, the (b) dimensionless radial displacement --- which is related to the hoop stress, and (c) the quantity $\bar{w}'$  --- which is related to the curvature, for the five mid-profiles at the buckling threshold when $\lambda=30$. The curves for the other values of $\lambda$ for each mid-profile overlap with the ones shown in this figures (not presented here).}
	\label{fig:ridgeCompareCone}
\end{figure*}

\rp{\subsubsection{Comparison to mirror buckling assumptions}\label{sec:comparison}}
To further investigate these deviations from our model and the implications for the mirror buckling assumption, we compare the numerical solution of the shallow shell equations to the analytical solution discussed in Section~\ref{SEC:PogorelovCone}. Figure~\ref{fig:ridgeCompareCone}(a) shows the ratio between the expected radial position of the ridge for the given indentation $\Delta_c$, i.e. $\rho_d = \Delta_c/(c_r \hat{Z})$, and the radial position computed by solving the shallow shell equations at the circumferential buckling threshold, $\rho_d^{num}$. All geometries, except spherical ($\bar{\kappa}_G=1$), are very close to 1, showing good agreement with the analytical model. In the spherical case, the ridge is consistently $10\%$ closer to the apex for all values of $\lambda$ investigated. 

Considering representative cases with $\lambda=30$ (but all values of $\lambda$ show the same behaviour), in Figs~\ref{fig:ridgeCompareCone}(b) and (c) we plot the dimensionless radial displacement $\bar{u}$ (related to the hoop stress) and the quantity $\bar{w}'$ (related to the curvature) \cite{Knoche2014}, at the circumferential buckling threshold, as a function of the dimensionless radial position measured from the centreline of the ridge, $\bar{s}$. From the dimensionless quantities in Eqns~\eqref{EQ:nondimensionalization_general}, the quantities $\bar{u}$, $\bar{w}'$ and $\bar{s}$ are derived as
\begin{align}
	\bar{s} &= \left(\frac{\lambda\hat{Z}}{(1-\nu^2)^{-1/4}\Delta_c}\right)\left(\rho-\rho_d\right), &&
	\bar{w}'= \left(\frac{(1-\nu^2)^{1/4}\Delta_c^{1/2}}{\hat{Z}^2\tilde{c}_r^{1/2}\lambda}\right)\frac{\upd^2 W}{\upd \rho^2},\\
	\bar{u} &= \left(\frac{\lambda (1-\nu^2)^{3/4}}{\Delta_c^{1/2}}\right)\frac{\upd^2 \Phi}{\upd \rho^2}. \nonumber
\end{align}
The solution obtained using the approach discussed in Section~\ref{sec:ridgeStructure} is shown in black in Fig.~\ref{fig:ridgeCompareCone}. Excluding the spherical case, the position of the ridge is close to the predicted position under the mirror buckling assumption. The radial distribution of the hoop stress and curvature is also well described by the analytical solution. For the spherical case ($\bar{\kappa}_G=1$), the profiles are consistent with the analytical solution in terms of the extreme values and qualitative radial behaviour. However, there is a shift in the radial location of the ridge towards the indentation point, as seen by the offset between the apex of the ridge and $\bar{s}=0$ in Fig.~\ref{fig:ridgeCompareCone}(b) and (c). 

There is a marked difference in the behaviour of the cases with $\bar{\kappa}_G=0,0.25,0.5,0.75$, and the spherical case $\bar{\kappa}_G=1$. For the non-spherical shells, all the the quantities in Fig.~\ref{fig:K_effect_scaled}, and the radial profiles in Fig.~\ref{fig:ridgeCompareCone}(b) and (c), show that the circular ridge buckles circumferentially in agreement with the mirror buckling assumption. Note that all these mid-profiles are \emph{conical} at the apex, i.e. the second derivative of the mid-profile in the radial direction is zero. This could be another reason why their behaviour follows the mirror buckling assumption. The deformed axisymmetric shape for small indentation is the same as a cone in all cases and this mirror buckled form may be largely maintained when the ridge moves to regions where the shell is no longer conical. 

The description, using the mirror buckling assumption, is valid in a qualitative sense for the spherical case; all the quantities show qualitative agreement with the mirror buckling predictions, but the offset in the radial position of the ridge and the difference in the rescaled hoop stress could be a reason why the mode number $m_c=3$ is observed instead of $m_c=5$, which appears to correspond to mirror buckled shells (such as deflating a spherical cap \cite{Knoche2014}).

In the description of the ridge in Sections \ref{sec:localSlope} and \ref{sec:bucklingOfPRidge} the local slope of the mid-profile is the governing parameter since mirror buckling is assumed. However, we showed that the emergence of mirror buckling depends on the Gaussian curvature and the mid-profile of the shells, when the outer boundary is clamped. Sun and Paulose \cite{Sun2021} discuss the role of principal curvatures in the definition of the length scales that arise from balancing bending and stretching, and how these affect the deformation of indented ellipsoidal and cylindrical shells. We considered the size of the planform as the characteristic length scale of the problem. If, instead, we rescale the problem using the shell thickness, $t$, we can investigate the scenario where the deformation induced by the indentation is localised in a small neighbourhood of the apex far away from the boundary (all the details are not shown in this work but they follow the exact same procedure with boundary conditions that impose decay of the solution for the radial coordinate going to infinity). In this `infinite' case a cone buckles with $\delta_c \approx 14 t$ and $m_c=3$, the same as the indented sphere \cite{Vella2011}. This reveals the importance of the boundary conditions, and their interaction with the shell mid-profile, in determining the circumferential buckling behaviour.

\rp{\subsubsection{Indentation force at the buckling threshold} \label{sec:indentationForce}}
\rp{It can be useful to understand how the indentation force at the buckling threshold is influenced by the shape of the shell as well as the implications of the mirror buckling assumption on the expected force. The dimensionless force is computed as the derivative, with respect to the dimensionless indentation depth, $\Delta$, of the dimensionless total strain energy, $\bar{\mathcal{E}}_{tot}$, including contribution of both the ridge energy $\bar{\mathcal{E}}_r$ and the mirror buckled inner dimple $\bar{\mathcal{E}}_{mb}$. We therefore write the dimensionless indentation force at the buckling threshold as:
	\begin{equation}
	\mathcal{F} = \frac{\partial \bar{\mathcal{E}}_{tot}}{\partial \Delta} = \frac{\partial \rho_d}{\partial \Delta}\frac{\partial \bar{\mathcal{E}}_{tot}}{\partial \rho_d}\label{eq:indentForce}
	\end{equation}
	with $\rho_d = \Delta c_r^{-1}\hat{Z}^{-1}$. Since $\hat{Z}$ and $c_r$ both depend only on $\rho_d$ we can write
	\begin{equation}
	\frac{\partial \rho_d}{\partial \Delta} = \left[\frac{\partial \Delta}{\partial \rho_d}\right]^{-1} = \left[\frac{\partial}{\partial \rho_d} \left(\rho_d c_r \hat{Z}\right) \right]^{-1}= \left[c_r \hat{Z} \left(1+\rho_d\left(c_r'/c_r +\hat{Z}'/\hat{Z}\right)\right)\right]^{-1} = \hat{Z}^{-1}
	\end{equation}
	where the prime indicates the derivative with respect to $\rho_d$. Somewhat surprisingly, this derivative does not depend on $c_r$.}

\begin{figure}
	\centering
	\includegraphics[width=0.5\textwidth]{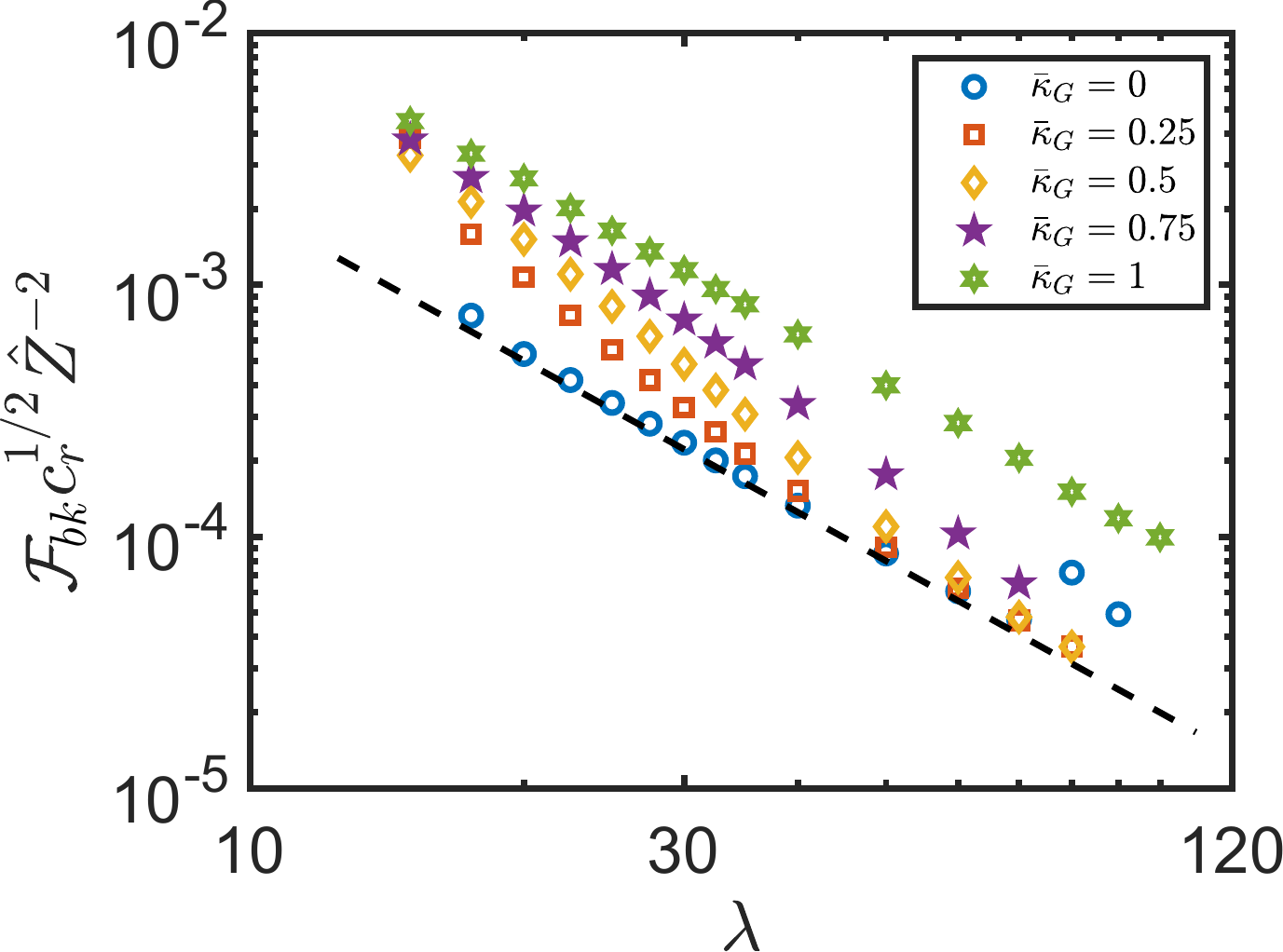}
	\caption{\rp{Dimensionless indentation force at the buckling threshold. The linear stability analysis results fit with the theoretical prediction for $\bar{\kappa}_G < 0.5$ and large $\lambda$. The results diverge for the remaining shapes to do the violation of the mirror buckling assumption.}}
	\label{fig:ForcePlot}
\end{figure}

\rp{The strain energy of the ridge was derived in Eqn~\ref{eq:energy}. We non-dimensionalise this energy, at the buckling threshold, using $E_n = \bar{\mathcal{E}} E t \mathcal{H}^4\mathcal{L}^{-2}$ leaving:
	\begin{equation}
	\bar{\mathcal{E}} \sim \lambda^{-3}\rho_d^{1/2}\hat{Z}^{5/2}.
	\end{equation}
	The dimensionless energy of the mirror buckled dimple, $\bar{\mathcal{E}}_{mb}$, is computed by generalising the expression proposed by Knoche and Kierfeld \cite{Knoche2014} and made dimensionless as
	\begin{equation}
	\mathcal{E}_{mb} \sim \lambda^{-2}\int_0^{\rho_d} \left(K_r^2+2\nu K_r K_{\theta} + K_{\theta}^2\right) 2 \pi  \rho \mathrm{d} \rho \label{eq:e_mb_integral}
	\end{equation}
	where the bending strains in the radial and circumferential directions, $K_r = 2 \bar{\kappa}_G \rho_d \hat{Z}^{-1}$ and $K_{\theta}= 2\hat{Z} \rho_d^{-1}$ are twice the respective curvatures of the underformed mid-profile since we have assumed mirror buckling. Noting that the dimple area is $\pi \rho_d^2$, the three terms of the integral scale as
	\begin{align}
		\int_0^{\rho_d} K_r^2\rho \mathrm{d} \rho \sim \kappa_G^2 \rho_d^{4} \hat{Z}^{-2} &&
		\int_0^{\rho_d} K_r K_\theta \rho \mathrm{d} \rho \sim \bar{\kappa}_G \rho_d^2 &&
		\int_0^{\rho_d}K_\theta^2\rho \mathrm{d} \rho \sim \hat{Z}^{2}
	\end{align}
}
\rp{The force at any indentation,  under the assumption of mirror buckling, is obtained using Eqn~\ref{eq:indentForce}. The force contribution of the ridge reads
	\begin{equation}
	\mathcal{F}_r \sim \frac{1}{2}\lambda^{-3}\rho_d^{-1/2}\hat{Z}^{3/2} + \frac{5}{2}\lambda^{-3}\rho_d^{1/2}\hat{Z}^{1/2}\hat{Z}'\label{eq:F_r}
	\end{equation}
	where $\hat{Z}'$ is the derivative of the local slope $\hat{Z}$ with respect to $\rho_d$, ie. the radial curvature of the ridge. The force contribution of the dimple reads
	\begin{equation}
	\mathcal{F}_{mb} \sim C_1 \kappa_G^2 \lambda^{-2} \hat{Z}^{-1}\rho_d^3 \left(2 \hat{Z}-\rho_d \hat{Z}'\right)+C_2\lambda^{-2} \kappa_G \rho_d \hat{Z}^{-1}+C_3\lambda^{-2}\hat{Z}' \label{eq:F_mb}
	\end{equation}}
\rp{The constants $C_i$ can be obtained by completing the integral in Eqn~\ref{eq:e_mb_integral} for a particular shell of revolution. However, we can make a general statement by noting that at the buckling threshold, $\rho_d \sim c_r\lambda^{-2}\hat{Z}^{-1}$, two contributions scale as $\lambda^{-2}$, the first term in Eqn~\ref{eq:F_r} and the third term in Eqn~\ref{eq:F_mb}, while all other terms decrease faster for increasing $\lambda$. From Eqn~\ref{eq:kG_shallow}, we also note that $\hat{Z}^2 \gg \hat{Z}'$, thus the ridge contribution dominates and we expect the force at buckling to scale as
	\begin{equation}\label{eq:ForceScaling}
	\mathcal{F}_{bk} \sim c_r^{-1/2}\lambda^{-2} \hat{Z}^{2}
	\end{equation}
	for large $\lambda$. This is plotted in Fig~\ref{fig:ForcePlot} for our range of shapes. For large $\lambda$, where the approximation given by \eqref{eq:ForceScaling} holds, the curves for $\kappa_G \leq 0.5$ collapse onto expected scaling, whereas there is some deviation for the other two cases. This deviation being due to violations of the mirror buckling assumption. The force at buckling is sensitive to variations from the mirror buckling assumption. In particular, the force analysis relies on $\hat{Z}$ (and $\hat{Z}'$) and $\rho_d$ based on the mirror buckling assumption. The ridge geometry, was compared to the mirror buckling assumption in Section~\ref{sec:comparison}. The analysis of the indentation force confirms the result that an indented spherical shell doesn't deform in a mirror buckled way, although the features of the ridge at the buckling threshold can be approximated, within a reasonable error, with the ones obtained using the mirror buckling assumption.}

\subsection{Finite Element Analysis}\label{sec:FEA}

Finite element analysis was used to validate the numerical results from the linear stability analysis on the shallow shell equations. The indentation of shallow shells of revolution was performed using the commercial finite element analysis package ABAQUS \cite{DassaultSystemes2018}. Four-node, doubly-curved, reduced integration thin shell elements (S4R) were used. A mesh sensitivity study was carried out to ensure the results were not significantly influenced by the mesh density. Nodes on the outer boundary were restrained in all degrees of freedom (fixed condition).  

\begin{figure}
	\centering
	\includegraphics[width=0.5\textwidth]{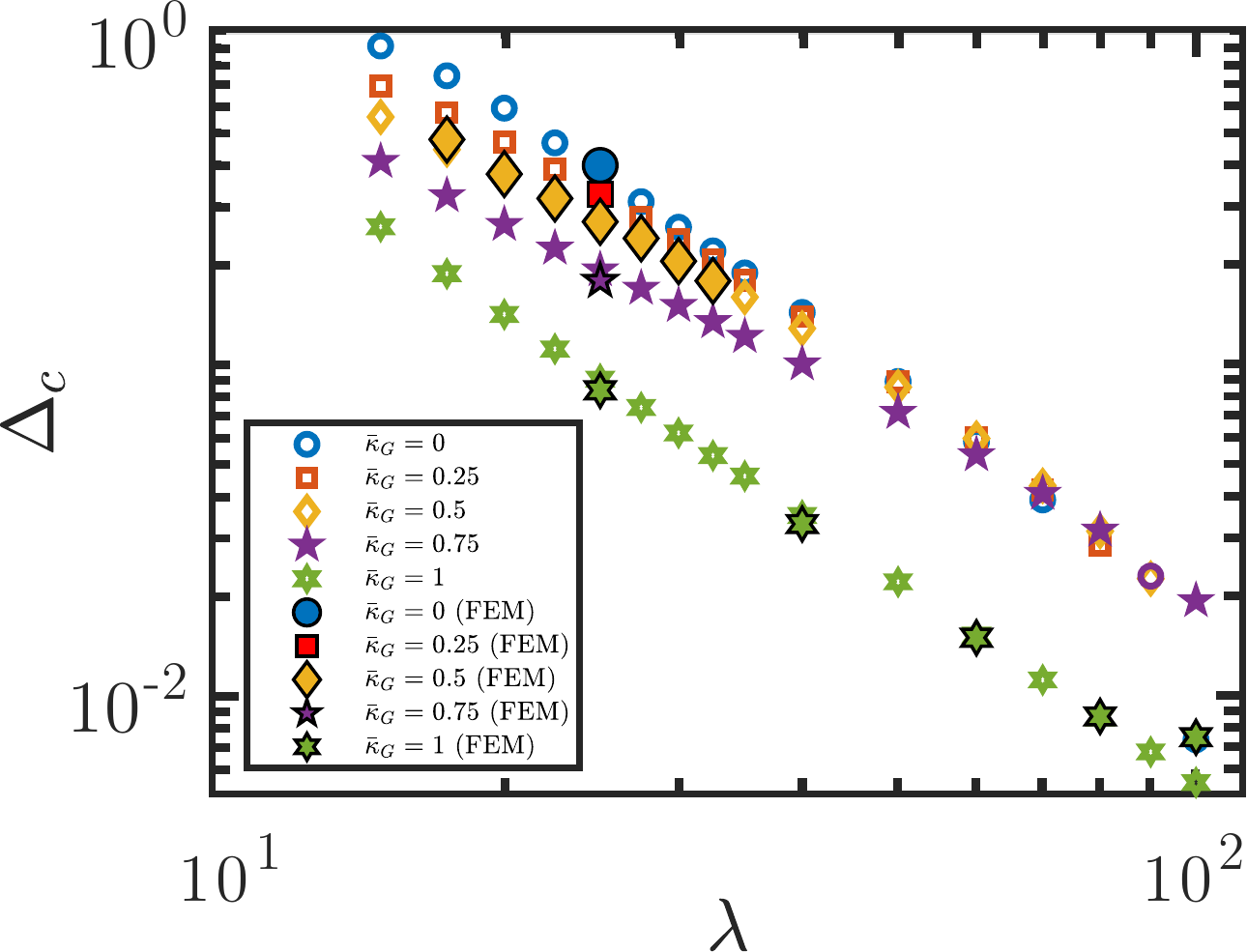}
	\caption{Dimensionless critical indentation depth $\Delta_c$ vs. $\lambda$ results from the linear stability analysis (open markers) and finite element analysis (filled markers). The results match very closely over the range of parameters considered. Comparison with Fig.~\ref{fig:K_effect_scaled} shows how the results collapse once the geometric parameter $c_r$ is included. }\label{fig:Indentation_Unscaled}
\end{figure}

A linear elastic material model was used with Young's modulus $E=1.5$~MPa and Poisson's ratio of 0.40. A non-linear static analysis was performed under displacement control of the apex point. To enable the solution to proceed through instabilities, a small amount of volume-proportional damping was added. This was reduced to the minimum required for the analysis to proceed, and always less than $5\%$ of the total strain energy. The results from the finite element and linear stability analyses, shown in Fig.~\ref{fig:Indentation_Unscaled}, are in very good agreement over the range of parameters considered.

\section{Conclusions}\label{sec:conclusion}

The earliest description for the shape of the axisymmetric ridge that forms in mirror buckled spherical shells is due to Pogorelov \cite{Pogorelov1988}. This formalism requires the ridge to connect an inner mirror buckled, unstretched, domain to an outer undeformed domain while also assuming the radial strain is negligible. In this paper we generalised this approach to characterise the ridge that forms in shells of revolution with non-negative Gaussian curvature deforming in the same, mirror buckled, fashion. We showed that the critical parameter controlling the ridge properties is the local (undeformed) slope at the ridge location. Under the additional constraint that the curvatures of the undeformed shape vary slowly with the radial coordinate, the variational problem to be solved for the ridge geometry in a generic shell of revolution can be recast in the same form proposed by Pogorelov \cite{Pogorelov1988}. 

We used the approach of Kierfield and Knoche \cite{Knoche2014} to estimate the relevant geometrical (width and curvature) and mechanical (hoop stress) features describing the ridge mechanics. These features were shown to only depend on the local slope of the undeformed mid-profile of the shell at the ridge location and the radial location of the ridge,  or the displacement of the apex (since they are univocally related assuming a mirror buckled shape). The scaling relations we derived are consistent with both the case of a cone \cite{Seffen2016a} and a sphere \cite{Knoche2014}. This allowed us to extend the result to a generic shell of revolution with non-negative Gaussian curvature. Comparison of these two limiting cases reveals that, although a common description for the ridge exists in dimensionless form, the behaviour of the ridge is markedly different when specialised to different geometries.

With the two-fold purpose to confirm our theoretical predictions and evaluate the mirror buckling assumption in indented shells of revolution, we analysed five mid-profiles with constant Gaussian curvature. This choice of shells allowed us to consider only one free parameter: the Gaussian curvature. However, our predictions can also be applied to profiles with varying Gaussian curvature and future work will  evaluate the mirror buckling assumption and circumferential buckling for these shells. Except for the spherical case, we found that all mid-profiles investigated were well-described by the our theoretical framework which assumes mirror buckling. \rp{The common feature of all the mid-profiles, except the spherical case, is the zero second derivative (with respect to the radial coordinate) at the apex; we can thus conjecture that the emergence of a mirror-buckled form is controlled by the shape at the apex.}


Our theoretical description shows that the geometrical and mechanical features of the axisymmetric ridge at onset of circumferential buckling are controlled by the local slope of the undeformed mid-profile at the ridge location and the shell geometry, through a dimensionless parameter $\lambda$. However, the phenomenology leading to different buckling modes remains uncertain. In addition to providing greater insight into the mechanics of thin shells, the results of our study could enable indentation to be used as a means to measure the mechanical properties of a wider range of thin shells. They could also lead to the design of shells with specific mechanical behaviours.

\noindent
{\bf Acknowledgements.}
MT is a member of the Gruppo Nazionale di Fisica Matematica (GNFM) of the Istituto Nazionale di Alta Matematica (INdAM).




\appendix
\section{Raw Data}\label{AppendixA}
\begin{figure*}[h!] 
	\centering
	\includegraphics[width=0.9\textwidth]{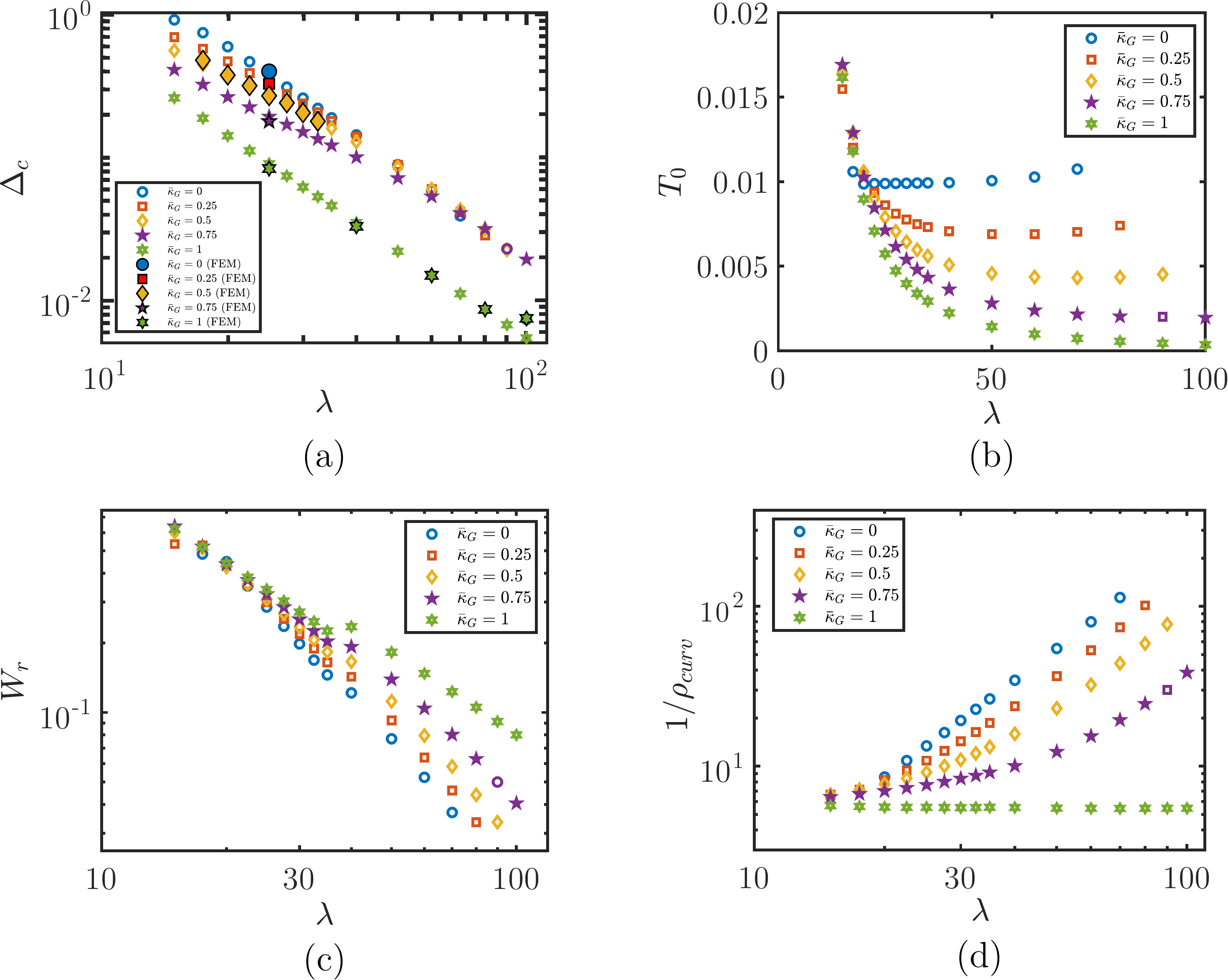}
	\caption{The influence of the (constant) Gaussian curvature of the undeformed shell on quantities at buckling as extracted from the linear stability analysis and the finite element simulations: (a) critical indentation, (b) maximum compressive hoop stress, (c) width of the ridge and (d) ridge curvature. Empty markers indicate the outcome of the linear stability analysis; filled markers indicate the outcome of the finite element simulations.}   \label{fig:K_effect}    
\end{figure*}

Figure~\ref{fig:K_effect} shows the raw data, as extracted from the numerical solution of the shallow shell equations, before they are rescaled using the relations in Eqns~\eqref{eq:buckling_scalings}. Figure~\ref{fig:K_effect}(a) shows the dimensionless critical indentation, the results from finite element analysis (see \ref{sec:FEA}) are shown as filled markers. There is excellent agreement between the results of the shallow shell analysis and the finite element analysis.  Figure~\ref{fig:K_effect}(b) shows the maximum (compressive) hoop stress in the circular ridge. Figure~\ref{fig:K_effect}(c) shows the width of the circular ridge, and Figure~\ref{fig:K_effect}(d) shows the ridge curvature computed at the radial position $\rho_d$. 

\end{document}